\documentclass[peerreview,onecolumn,11pt,draftclsnofoot]{IEEEtran}%% bare_jrnl.tex
\usepackage{amsmath}%% for current contact information.
\usepackage{amsfonts}%%
\usepackage{epsfig}%% This is a skeleton file demonstrating the use of IEEEtran.cls
\usepackage{amssymb}%% (requires IEEEtran.cls version 1.7 or later) with an IEEE journal paper.
\usepackage{graphicx}%% Support sites:d
\usepackage{amssymb,amsmath}%% http://www.michaelshell.org/tex/ieeetran/
\usepackage{cite}%% http://www.ctan.org/tex-archive/macros/latex/contrib/IEEEtran/
\usepackage{color,soul}%% and
\begin{document}
\title{Robust Spectrum Sharing via Worst Case Approach}

\author{Saeedeh Parsaeefard, \IEEEmembership{Student Member, IEEE,} and
        Ahmad R. Sharafat, \IEEEmembership{Senior Member, IEEE}
        %and~Jane~Doe,~\IEEEmembership{Life~Fellow,~IEEE}% <-this % stops a space
\thanks{Manuscript received March 19, 2011. This work was supported in part by Tarbiat Modares University, and in part by Iran Telecommunications Research Center, Tehran, Iran, under PhD Research Grant TMU 88-11-124.

The authors are with the Department of Electrical and Computer
Engineering, Tarbiat Modares University, P. O. Box 14155-4838,
Tehran, Iran. Corresponding author is A.~R.~Sharafat (e-mail:
sharafat@modares.ac.ir).}}

\maketitle

\begin{abstract}
This paper considers non-cooperative and fully-distributed power-allocation for secondary-users (SUs) in spectrum-sharing environments when normalized-interference to each secondary-user is uncertain. We model each uncertain parameter by the sum of its nominal (estimated) value and a bounded additive error in a convex set, and show that the allocated power always converges to its equilibrium, called robust Nash equilibrium (RNE). In the case of a bounded and symmetric uncertainty set, we show that the power allocation problem for each SU is simplified, and can be solved in a distributed manner. We derive the conditions for RNE's uniqueness and for convergence of the distributed algorithm; and show that the total throughput (social utility) is less than that at NE when RNE is unique. We also show that for multiple RNEs, the the social utility may be higher at a RNE as compared to that at the corresponding NE, and demonstrate that this is caused by SUs' orthogonal utilization of bandwidth for increasing the social utility. Simulations confirm our analysis.

\end{abstract}

%\IEEEpeerreviewmaketitle
\section{Introduction}

\IEEEPARstart{O}{pportunistic} spectrum sharing by secondary users (SUs) in cognitive radio networks (CRNs) is a promising approach for improving spectrum efficiency in future wireless communication systems. In this setup, each SU aims to maximize its utility subject to its power limit and other regulatory restrictions, such as spectrum mask and/or the maximum amount of interference induced to licensed (primary) users (PUs) \cite{Haykin,Goldsmith}.

Because of the inherently decentralized and competitive nature of allocating power to SUs, game theory is an appropriate tool for analyzing such systems \cite{MIMOcognitivescutari,Robusthaykin,ScutariVI,ProbabilisticIWFA,Robustnew}. In this context, the power allocation problem is formulated as a strategic non-cooperative game, where each SU is a player that competes with others by choosing a transmission strategy for maximizing its own utility, defined as its throughput. In a game, Nash equilibrium (NE) is a state, consisting of the strategy space and utility values, at which there is no incentive for any player to change its strategy, provided that other players' strategies are not changed. As such, one needs to derive the conditions for existence and uniqueness of NE, to develop a distributed algorithm for reaching NE, and to examine the convergence conditions for that algorithm.

The well known iterative water-filling algorithm (IWFA) for
reaching NE of a power allocation game, as well as the conditions for existence of NE and convergence conditions of IWFA are proposed in \cite{Yu}. Besides, in \cite{MIMOcognitivescutari, ScutariVI, scutari5,Luo1, Luo2, Nash1, Nash2,SimultanousWFA,Spectrumsharing}, other distributed water-filling based algorithms and conditions for their convergence are studied, and the sufficient conditions for existence and uniqueness of NE under different power constraints and system models are derived. In many of the existing power allocation schemes, it is assumed that the channel state information, and interference from other SUs plus noise (IN) are available to each SU. However, this assumption may not be valid in practice, due to the time-varying environment and inaccurate measurements, resulting in uncertainties in parameter values.

Motivated by the aforementioned challenge, we wish to
develop a robust game-theoretic approach for tackling such
uncertainties in the power allocation problem. In doing so, we model each uncertain parameter by the sum of its nominal (estimated) value plus a bounded additive error, the collection of which form the uncertainty region. The corresponding game is based on the robust optimization theory \cite{Robustgame}, where each SU obtains its transmit power level that maximizes its throughput in the worst case instance in the uncertainty region \cite{selecectedrobust}, and the equilibrium of such a game is called the robust Nash equilibrium (RNE). We will show that when the uncertainty region is a closed convex set, RNE always exists, but a closed form solution for RNE may not be obtainable for some forms of the uncertainty region. We focus on bounded symmetric uncertainties, and utilize the framework in \cite{Nash1,Nash2} to obtain the necessary conditions for uniqueness of RNE. Moreover, to reach RNE, we will develop simultaneous as well as sequential distributed power allocation algorithms, and obtain the conditions for their convergence.

Another important subject in this paper is to determine how
uncertainty affects the total throughout, i.e., the social utility of SUs at RNE as compared to that of a game with complete information. We will show that when RNE is unique, uncertainty reduces the total throughput of SUs, which is not the case for multiple NEs.  When multiuser interference is high, multiple NEs may exist in IWFA-based algorithms. In such cases, uncertainty may lead to a more orthogonal power allocation at RNE, resulting in a higher total throughput of SUs as compared to the case of a game with complete information.

Recent works that assume uncertainty in parameter values include \cite{ProbabilisticIWFA,Robustnew}. In
\cite{ProbabilisticIWFA}, a probabilistic robust IWFA is proposed, where IN levels are uncertain. By assuming a uniform probability density function (pdf) for uncertainty in IN levels, the power allocation problem is converted to the conventional IWFA, but the IN level is multiplied by a factor that corresponds to the stochastic nature of uncertainty. In \cite{Robustnew}, the worst case robust optimization theory is used when uncertainty in the channel state information between users is bounded in an ellipsoid region to derive the conditions for NE's existence and uniqueness, and a distributed algorithm that needs additional message passing between users is proposed. Our main contribution in this paper is a simplified robust distributed power allocation scheme that does not need any additional message passing in the system. In addition, to provide an insight into the performance of our robust power allocation scheme, we compare the social utility at its RNE with that of the conventional game at its NE. We do this for both unique and multiple RNE cases, and derive the conditions for RNE's uniqueness and for convergence of our distributed algorithms.

This paper is organized as follows. In Section II, we present the system model of a spectrum sharing environment, and formulate a robust game for the power allocation problem, when SUs channel state information and IN levels are uncertain. In Section III, we investigate the existence and uniqueness conditions of RNE in the proposed game, and study the effect of uncertainties on the total throughput at RNE as compared to NE of the complete information game. Our distributed algorithms are proposed in Section IV, follows by simulation results in Section V, and conclusions in Section VI.

\section{System Model and Problem Statement}
We consider a multiuser CRN with $\mathcal{M}=\{1,\cdots, M\}$ SUs sharing $\mathcal{K}=\{1,\cdots,K\}$ orthogonal narrow band sub-channels. The bandwidth of each sub-channel is much less than the coherence bandwidth of the wireless channel, meaning that the channel response of each sub-channel is flat. The transmit power vector of the $i^{\scriptsize{\textnormal{th}}}$ SU over all sub-channels is $\mathbf{p}_{_i}=[p^{1}_{i},\cdots,p^{K}_{i}]$, where $p^{k}_{i}$ is the transmit power of the $i^{\scriptsize{\textnormal{th}}}$ SU in the $k^{\scriptsize{\textnormal{th}}}$ sub-channel. The received signal at the corresponding receiver in sub-channel $k$ is
\begin{equation}\label{recive}
    r^{k}_{i}=\sum^{M}_{i=1} p^{k}_{i}h^{k}_{ji}+\sigma_{i}^{k}, \quad
    \forall i \in \mathcal{M} , \quad  k \in \mathcal{K},
\end{equation}
where $h^{k}_{ji}$ is the fading sub-channel gain from the $j^{\scriptsize{\textnormal{th}}}$ transmitter to the
$i^{\scriptsize{\textnormal{th}}}$ receiver on the
$k^{\scriptsize{\textnormal{th}}}$ sub-channel, and
$\sigma_{i}^{k}$ is the noise power in the $k^{\scriptsize{\textnormal{th}}}$ sub-channel of the
$i^{\scriptsize{\textnormal{th}}}$ SU. At receiver $i$, signal-to-interference-plus-noise-ratio (SINR) in sub-channel $k$ is
%\begin{equation}\label{SINR}
      $\gamma^{k}_{i}= \frac{p^{k}_{i}}{s^{k}_{i}}$,
%\end{equation}
where $s^{k}_{i}=\frac{\sum_{i\neq j}
p^{k}_{j}h^{k}_{ji}+\sigma_{i}^{k}}{h_{ii}^{k}}$. Hence, for each
user, the achievable rate for the
$i^{{\scriptsize{\textnormal{th}}}}$ receiver is obtained by
%\begin{equation}\label{Rate1}
$R_i= \sum_{k=1}^{K}\log(1+ \gamma^{k}_{i})$. We assume that $h^{k}_{ii}$ and noise-plus-interference generated by other SUs are estimated by the $i^{\scriptsize{\textnormal{th}}}$ receiver. The receiver calculates the value of $s_{i}^{k}$ in each sub-channel, and sends it to the corresponding transmitter via the feedback channel. The transmit power of each SU is subject to the following constraints:
\begin{description}
    \item [$C_1$:] The total transmit power of the $i^{\scriptsize{\textnormal{th}}}$ SU over all sub-channels is limited by its maximum power budget, i.e.,
%\begin{equation}\label{powerbudget}
    $\sum_{k=1}^{K} p^{k}_{i}\leq
    p_{i}^{\scriptsize{\textnormal{max}}}$.
%\end{equation}
    \item [$C_2$:]The transmit power of each SU on each sub-channel is limited, i.e.,
%\begin{equation}\label{powermask}
$0\leq p^{k}_{i}\leq p^{k}_{\scriptsize{\textnormal{mask}}}$,
%\end{equation}
where $p^{k}_{\scriptsize{\textnormal{mask}}}$ is the spectral mask on sub-channel $k$.
\end{description}
We assume that users cannot perform interference cancelation.

\subsection{Game Formulation}
In a noncooperative CRN, each SU aims to maximize its own
utility, defined as its throughput, subject to $C_1$ and $C_2$. Such a greedy behavior can be analyzed by game theory, where each SU is a player with a set of power allocation strategies over $K$ sub-channels, defined as
\begin{equation}\label{Powerstratgy}
    \mathcal{P}_{i} \triangleq \{ \mathbf{p}_{_i} |  \sum_{k=1}^{K} p^{k}_{i} \leq  p_{i}^{\scriptsize{\textnormal{max}}}, \quad 0\leq p^{k}_{i}\leq  p^{k}_{\scriptsize{\textnormal{mask}}}, \quad  \forall k \in \mathcal{K}\}.
\end{equation}
This game is denoted by $\mathcal{G} \triangleq
\langle \mathcal{M},\{ \mathcal{P}_{i}\}_{i \in \mathcal{M}},
\{u_{i}(\mathbf{p})\}_{i\in \mathcal{M}}\rangle $, where $u_{i}(\mathbf{p})$ is the utility of user $i$ that depends on the chosen strategy vector of all users  $\mathbf{p}=[\mathbf{p}_{_i},\mathbf{p}_{_{-i}}]$, where
$\mathbf{p}_{_{-i}}=[\mathbf{p}_{_1},\cdots,\mathbf{p}_{_{i-1}},\mathbf{p}_{_{i+1}},\cdots,\mathbf{p}_{_M}]$. The optimal strategy for the
$i^{{\scriptsize{\textnormal{th}}}}$ SU, given the transmit power levels of other SUs, is the solution to the following
problem
\begin{eqnarray}\label{IWFutility1}
  \max_{\mathbf{p}_{_i}, \mathbf{p}_{_{-i}}} & \sum_{k=1}^{K}\log(1+\frac{p^{k}_{i}}{s^{k}_{i}})
 \end{eqnarray}
\[\mathrm{subject~to} \left\{
\begin{array}{l l}
 C_1: \sum_{k=1}^{K} p^{k}_{i} \leq  p_{i}^{\scriptsize{\textnormal{max}}} \nonumber \\
C_2: 0\leq p^{k}_{i}\leq  p^{k}_{\scriptsize{\textnormal{mask}}} \nonumber \\
\end{array} \right. \]
The solution to (\ref{IWFutility1}), obtained by IWFA is
\begin{equation}\label{IWFA}
   p_{i}^{k}=[\frac{1}{\lambda_{i}}-s^{k}_{i}]^{p^{k}_{\scriptsize{\textnormal{mask}}}}_{0},
\end{equation}
where $[x]^{b}_{a}$  for $a<b$ denotes the Euclidean projection
of $x$ onto the interval $[a,b]$, i.e., $[x]^{b}_{a}=a$ if $x<a$, $[x]^{b}_{a}=ax$ if $a<x<b$, and $[x]^{b}_{a}=b$ if $b<x$. The parameter $\lambda_{i}$ is the Lagrange multiplier that satisfies $C_1$ for the $i^{{\scriptsize{\textnormal{th}}}}$ user, i.e.,
\begin{eqnarray}\label{optimaloptlambda}
\lambda_{i} \times (\sum^{K}_{k=1}p_{i}^{k}-
p_{i}^{\scriptsize{\textnormal{max}}})=0,  \qquad\forall i \in
\mathcal{M}.
\end{eqnarray}
The strategy profile
$\mathbf{p}^{*}=\{\mathbf{p}^{*}_{_1},\cdots,\mathbf{p}^{*}_{_M}\}$ is NE for the game $\mathcal{G}$ if
\begin{equation}\label{NEpoint}
 u_{i}(\mathbf{p}^{*}_{_i},\mathbf{p}^{*}_{_{-i}})\geq
 u_{i}(\mathbf{p}_{_i},\mathbf{p}^{*}_{_{-i}}), \quad \forall \mathbf{p}_{_{i}} \in \mathcal{P}_i, \quad \forall i \in \mathcal{M}.
\end{equation}
Since the utility function $u_{i}(\mathbf{p})$ is concave on $\mathbf{p}_{_i}$, the game $\mathcal{G}$ has a nonempty solution set for any set of channels, spectral mask constraints, and transmit power levels \cite{NEexistence}. However, uniqueness of NE depends on cross channel gains between SUs, and different conditions for uniqueness of NE in the game $\mathcal{G}$ are proposed in \cite{Nash1}. Generally, if multiuser interference is low, the game has a unique NE, and when multiuser interference is high, the game has multiple NEs.

\subsection{Robust Counterpart Game Formulation}
In the above algorithm, it is assumed that the exact value of $s_{i}^{k}$ for each subchannel is available to the respective SU receiver with no error. This value is transmitted to the corresponding SU transmitter via the feedback channel. However, due to the dynamic nature of CRNs manifested in channel variations, users' movements, new users in the system, as well as the delay in the feedback channel, errors are introduced in $s_{i}^{k}$, which invalidate the assumption that the error-free value of $s_{i}^{k}$ is available to the the respective transmitter. This means that power allocation based on (\ref{IWFA}) cannot guarantee the optimum utility of users in reality.

To tackle this problem, in our formulation, players assume that each parameter in the transmitter is uncertain, modeled by a deterministic nominal (estimated) value plus an uncertain term, i.e.,
%\begin{eqnarray}\label{symetricuncertainity}
$s^{k}_{i} = \bar{s}^{k}_{i}+\hat{s}^{k}_{i}$,
%\end{eqnarray}
where $\bar{s}^{k}_{i}$ and $\hat{s}^{k}_{i}$ are the nominal
value and error in  $s^{k}_{i}$ for the $i^{\text{th}}$
transmitter, respectively. Similar to the robust optimization in game theory \cite{Robustgame}, we assume that $s^{k}_{i}$
statistics are unknown, but its variations are bounded, i.e., deviations from their nominal values are bounded to  $\mathcal{R}_{s}$,
 \begin{equation}
 \hat{\mathbf{s}}_{i}=(\mathbf{s}_{i}-\bar{\mathbf{s}}_{i}) \in \mathcal{R}_{s} , \quad
  \forall  i \in \mathcal{M},\quad \forall  k \in \mathcal{K}.
 \end{equation}
where $\mathbf{s}_{i}=[s_{i}^{1},\cdots,s_{i}^{K}]$, $\hat{\mathbf{s}}_{i}=[\hat{s}_{i}^{1},\cdots,\hat{s}_{i}^{K}]$, and $\bar{\mathbf{s}}_{i}=[\bar{s}_{i}^{1},\cdots,\bar{s}_{i}^{K}]$.
The uncertain parameter is a new variable in the optimization problem of each user, and the robust counterpart of (\ref{IWFutility1}) is changed \cite{selecectedrobust} to
\begin{equation}\label{IWFrobustcounterpart}
  \max_{\mathbf{p}_{i}, \mathbf{p}_{-i}} \min_{s^{k}_{i}\in \mathcal{R}_{s} }\sum_{k=1}^{k}\log(1+\frac{p^{k}_{i}}{s^{k}_{i}})
 \end{equation}
 \[\mathrm{subject~to} \left\{
\begin{array}{l l}
 C_1: \sum_{k=1}^{K} p^{k}_{i} \leq  p_{i}^{\scriptsize{\textnormal{max}}} \nonumber \\
 C_2: 0 \leq p^{k}_{i}\leq  p^{k}_{\scriptsize{\textnormal{mask}}} \nonumber \\
\end{array} \right. \]
We define the corresponding robust game for the objective function (\ref{IWFrobustcounterpart}) as $\widetilde{\mathcal{G}}
\triangleq \langle \mathcal{M},\{\mathcal{P}_{i}\}_{i \in \mathcal{M}}, \{\widetilde{u}_{i}\}_{i\in\mathcal{M}} \rangle$, where for the optimal solution to (\ref{IWFrobustcounterpart}), the utility of
$\widetilde{\mathcal{G}}$ is
$\widetilde{u}_{i}(\mathbf{p}_{_{i}},\mathbf{p}_{_{-i}},\mathcal{R}_{s})$.
For both games $\mathcal{G}$ and $\widetilde{\mathcal{G}}$, the sets $\mathcal{M}$ and $\mathcal{P}_{i}$ are the set of all SUs
and the set of strategy profile of each SU from
(\ref{Powerstratgy}), respectively. In this case, the strategy
profile $\{\widetilde{\mathbf{p}}^{*}_{_{1}},\cdots,\widetilde{\mathbf{p}}^{*}_{_{M}}\}$ is RNE for the game
$\widetilde{\mathcal{G}}$ if
\begin{equation}\label{NEpoint}
 \widetilde{u}_{i}(\widetilde{\mathbf{p}}^{*}_{_{i}},\widetilde{\mathbf{p}}^{*}_{_{-i}},\mathcal{R}_{s})\geq
 \widetilde{u}_{i}(\mathbf{p}_{_{i}},\widetilde{\mathbf{p}}^{*}_{_{-i}},\mathcal{R}_{s}),\; \forall \mathbf{p}_{_{i}} \in \mathcal{P}_{i}, \;\forall s^{k}_{i}\in \mathcal{R}_{s}, \; \forall i \in \mathcal{M}.
\end{equation}

\section{Analysis of RNE}
We now present our analysis on RNE's existence and uniqueness. Recall that by considering uncertainty in the game, equilibrium analysis for $\widetilde{\mathcal{G}}$ is more complicated than that for $\mathcal{G}$. This is because the strategy of each user depends on strategies of other users as well as on users' uncertainty regions. In addition, by considering uncertainty in the system, a new coupling is introduced in the game, which requires new signalling between users \cite{Robustnew}. Hence, design and implementation of such a network is more complicated when RNE is considered. Besides, considering uncertainty in the system via worst case optimization theory is a conservative approach, as users assume the worst case conditions for errors, which may be unrealistic. Hence, comparing the total achieved utility of the system (i.e., social utility) at RNE with that at NE provides a measure of performance when robustness is introduced. Considering these issues, we wish to
\begin{itemize}
  \item simplify the robust game to the extent possible, with a view to simplifying the implementation of the system, and
  \item determine the relationship between the social utility at RNE and at NE in both cases of unique NE and multiple NEs.
\end{itemize}

\subsection{Existence and Uniqueness of RNE}

For any realization of error in the uncertainty region and strategy profile, the worst case utility function for $\widetilde{\mathcal{G}}$, denoted by $\psi_i(\mathbf{p}_{_{i}},\mathbf{p}_{_{-i}})$ is
\begin{equation}\label{rho}
    \psi_i(\mathbf{p}_{_{i}},\mathbf{p}_{_{-i}})\triangleq \min_{s^{k}_{i}\in \mathcal{R}_{s}
    }\sum_{k=1}^{k}\log(1+\frac{p^{k}_{i}}{s^{k}_{i}}), \quad
    \forall i \in \mathcal{M}.
\end{equation}
When $\mathcal{R}_{s}$ is bounded and convex,
$\psi_i(\mathbf{p}_{_{i}},\mathbf{p}_{_{-i}})$ is continuous, and is concave on $\mathbf{p}_{_{i}}$, when $\mathbf{p}_{_{-i}}$ is fixed. Therefore, for any channel realization, any bound on the transmit power, and any constraint on spectral mask, there is an equilibrium, called RNE, for the game $\widetilde{\mathcal{G}}$ (Theorem 2 in \cite{Robustgame}).

Although one can establish the existence of RNE in a straight forward manner from the characteristics of $\mathcal{R}_s$, obtaining RNE requires excessive calculations and depends on the representation of the uncertainty region, meaning that the optimal transmit power cannot be obtained in a closed form. Hence, the conditions for RNE's uniqueness cannot be obtained in general by the fixed point approach and contraction mapping as in \cite{Nash1,Nash2}. Nevertheless, for the game $\widetilde{\mathcal{G}}$ when $\hat{s}^{k}_{i}$  is symmetrically distributed in the uncertainty region, i.e.,
\begin{equation}\label{Uncertainityregionlinear}
\hat{s}^{k}_{i}=[-\varepsilon^{k}_{i}\bar{s}^{k}_{i},\varepsilon^{k}_{i}
\bar{s}^{k}_{i}],
\end{equation}
we will obtain such conditions. Note that this form of the uncertainty region is valid for modeling statistical uncertainty in independent parameters.

\textbf{Proposition 1.} If (\ref{Uncertainityregionlinear}) holds, RNE of the robust game $\widetilde{\mathcal{G}}$ is the same as NE of the game $\mathcal{G}$ with the same number of users and strategy profile, and the optimal strategy is the solution to the following problem
\begin{equation}\label{utilitywithuncertainties}
\max_{\mathbf{p}_{_{i}}, \mathbf{p}_{_{-i}}}
\sum_{k=1}^{k}\log(1+\frac{p^{k}_{i}}{\bar{s}^{k}_{i}(1+\varepsilon^{k}_{i})})
\end{equation}
\[\mathrm{subject~to} \left\{
\begin{array}{l l}
C_1: \sum_{k=1}^{K} p^{k}_{i} \leq  p_{i}^{\scriptsize{\textnormal{max}}} \nonumber \\
C_2: p^{k}_{i}\leq  p^{k}_{\scriptsize{\textnormal{mask}}}. \nonumber \\
\end{array} \right. \]
\begin{proof}
See Appendix A.
\end{proof}
By using Proposition 1, uncertainty in system parameters can be modeled by the utility function in a deterministic manner.
Comparing the game $\mathcal{G}$ and the robust game $\widetilde{\mathcal{G}}$ whose utility is (\ref{utilitywithuncertainties}), we note that the robust game $\widetilde{\mathcal{G}}$ is changed to the conventional IWFA, and system parameters depend only on the uncertainty region. According to the Karush-Kuhn-Tucker (KKT) conditions \cite{boydconvexbook}, the optimal solution to (\ref{utilitywithuncertainties}) can be obtained by the Lagrange dual function as
\begin{equation}\label{optimalopt1robustgame}
    p_{i}^{k}=[\frac{1}{\lambda_{i}}-s_{i}^{k}(1+\varepsilon_{i}^{k})]_{0}^{p^{k}_{\scriptsize{\textnormal{mask}}}},
\end{equation}
where $\lambda_i$ is the
nonnegative Lagrange multiplier that satisfies
(\ref{optimaloptlambda}).

Next we derive the uniqueness condition of RNE for robust game using the framework in \cite{Nash1}.

\textbf{Proposition 2.} When (\ref{Uncertainityregionlinear}) holds, RNE is unique if
\begin{equation}\label{Proposition 2.1}
   \!\! \min\{\frac{\rho(\bar{\textbf{S}}(k)+\bar{\textbf{S}}^{T}(k))}{2},\|\bar{\textbf{S}}(k)\|_2\}+\|\mathbf{s}(k)\|_2<1, \quad \forall k \in \mathcal{K},
\end{equation}
where $\bar{\textbf{S}}(k)$ is a $M \times M$ matrix whose elements are
\begin{eqnarray}\label{Proposition 2.3}
 \bar{S}_{ij}(k)\triangleq \left\{\begin{array}{l l}
0 \qquad\qquad \scriptsize{\textnormal{if}}
\qquad i=j \\
 \frac{\bar{h}^{k}_{ji}}{\bar{h}^{k}_{ii}} \qquad\quad\, \scriptsize{\textnormal{if}} \qquad i\neq j, \end{array} \right.
 \end{eqnarray}
the value of $\|\bar{\textbf{S}}(k)\|_2$ is the $l_2$-norm of $\bar{\textbf{S}}(k)$, and $\mathbf{s}(k)=[\varepsilon^{k}_1,\cdots,\varepsilon^{k}_M]$. When $\bar{\textbf{S}}(k)$ is symmetric, (\ref{Proposition 2.1}) reduces to
\begin{equation}\label{Proposition 2.2}
    \rho(\bar{\textbf{S}}(k))+\|\mathbf{s}(k)\|_2<1, \qquad k \in \mathcal{K}.
\end{equation}
\begin{proof}
See Appendix B.
\end{proof}
Proposition 2 indicates that when interference is low, e.g., when cross channel gains between SUs are less than the direct channel gain, RNE is unique.

\subsection{Comparison of Social Utility at RNE and at NE}

Now we discuss the effect of uncertainty on RNE of
$\widetilde{\mathcal{G}}$ as compared to NE of $\mathcal{G}$ in
terms of the total throughput of SUs and the number of
sub-channels utilized by each SU.

\textbf{Lemma 1.} When Proposition 2 holds, the social utility at RNE of $\widetilde{\mathcal{G}}$ is less than that of
$\mathcal{G}$.
\begin{proof}
See Appendix C.
\end{proof}

Based on Lemma 1, uncertainty definitely reduces the total
throughput of SUs when RNE is unique. This may not be
true when interference is high, i.e., when we encounter multiple NEs in both $\mathcal{G}$ and $\widetilde{\mathcal{G}}$. By considering robustness in such cases, we may see a higher total throughput for a particular RNE as compared to the case for the corresponding NE, depending on the interfering and direct channel gains between SUs \cite{ProbabilisticIWFA,Robustnew}.
To explain the effect of uncertainty in the case of multiple NEs, we define the orthogonal equilibrium in interference channels. At orthogonal equilibrium, the transmit power levels of different users over each sub-channel are non-overlapping, meaning that only one transmitter uses a given sub-channel.

\textbf{Lemma 2.} Uncertainty in the game $\widetilde{\mathcal{G}}$ causes convergence to a RNE that has more orthogonality than NE of the game $\mathcal{G}$.

\begin{proof}
See Appendix D.
\end{proof}

Note that when interference is high, which corresponds to multiple NEs, the IWFA is suboptimal, but orthogonal power allocation is optimal \cite{SimultanousWFA,Spectrumsharing}. In such cases, SUs are forced to the orthogonal power allocation, where the total throughput at RNE may be higher than that of the corresponding NE, depending on the values of uncertain parameters, i.e., the interfering and direct channel gains between SUs, the amount of noise, and power limitations \cite{ProbabilisticIWFA,Robustnew}. When uncertainty is considered, the social utility of SUs is a non-smooth and non-convex function, and hence, in general, one cannot state that its value at RNE is higher or lower than its corresponding value at NE of the conventional game with complete information (no uncertainty).

\section{Distributed Algorithms}
A distributed power allocation algorithm is very desirable for
spectrum sharing environments. In order to develop such algorithms, we employ an iterative scheme that is based on the best response (\ref{optimalopt1robustgame}), similar to  the conventional distributed IWFA \cite{Nash2}. Generally, There are two classes of iterative algorithms: sequential algorithms in which users sequentially update their strategies according to a given schedule; and simultaneous algorithms in which all users update their strategies at the same time. In what follows, we describe these two classes.

\begin{table}[h]
\begin{tabular}{l}
\textbf{\quad\quad\quad Simultaneous Distributed Robust Worst Case IWFA}\\
\hline
\\\quad\quad For $t=0$, set any feasible power allocation $\textbf{p}_i(0) $ for all $i \in \mathcal{M}$.
\\ % Entering row contents
\quad \quad\quad  For $t=1,2,\cdots,T$:   \\ \quad\quad\quad\quad\quad
Calculate $\textbf{p}_i(t) $ from (\ref{optimalopt1robustgame}) ,
\,\,\, $\forall i \in  \mathcal{M}$.
\\\quad  \quad\quad End. \\[1ex]
\hline
\end{tabular}
\end{table}
\begin{table}[h]
%\caption{Simultaneous distributed worst case robust IWFA} %title of the table
%\centering % centering table
\begin{tabular}{l}
\textbf{\quad\quad\quad Sequential Distributed Robust Worst Case IWFA}\\
\hline
\quad\quad For $t=0$, set any feasible power allocation $\textbf{p}_i(0) $ for all $i \in \mathcal{M}$.\\
\quad \quad\quad  For $t=1,2,\cdots,T$:   \\
\quad \quad\quad  $\forall \,i \in \mathcal{M}$, consider $w=\text{mod} (t,M)$, \\
\quad\quad\quad\quad\quad if $i=w$, calculate $\textbf{p}_i(t) $ from
(\ref{optimalopt1robustgame}), \,\,\, $\forall i \in
\mathcal{M}$. \\
\quad\quad\quad\quad\quad Otherwise, $\textbf{p}_i(t+1)=
\textbf{p}_i(t)$.
\\\quad  \quad\quad End. \\[1ex] % [1ex] adds vertical space
\hline
\end{tabular}
\end{table}

\textbf{Proposition 3.} The distributed algorithm converges to the unique RNE from any initial power allocation $\textbf{p}_i(0)$ if
%\begin{equation}\label{condition distributed algorithm.1}
$\|\mathbf{\bar{S}}^{\scriptsize{\textnormal{max}}}\|_{2}+\sqrt{|\mathcal{M}|}\|\mathbf{s}^{\scriptsize{\textnormal{max}}}\|_{2}
< 1$,
%\end{equation}
where $\mathbf{\bar{S}}^{\scriptsize{\textnormal{max}}}$ is a $M \times M$ matrix, whose elements are
\begin{eqnarray}\label{TT}
 \bar{S}_{ij}^{\scriptsize{\textnormal{max}}} \triangleq  \left\{\begin{array}{l l}
0 \qquad\qquad\qquad\,\; \scriptsize{\textnormal{if}}\qquad i=j \\
 \max_{k \in \mathcal{K}}\frac{\bar{h}^{k}_{ji}}{\bar{h}^{k}_{ii}} \qquad \scriptsize{\textnormal{if}} \qquad i\neq j, \end{array} \right.
 \end{eqnarray}
 and $\mathbf{s}^{\scriptsize{\textnormal{max}}}$ is a $M \times 1$ vector whose $i^{\scriptsize{\textnormal{th}}}$ element is $\max_{k \in \mathcal{K}} \bar{s}_{i}^{k}\varepsilon_{i}^{k} $
\begin{proof}
See Appendix E.
\end{proof}

\section{Simulation Results}
Now we provide simulation results to get an insight into the
performance of $\widetilde{\mathcal{G}}$ for different bounds on uncertainty as compared to $\mathcal{G}$.

\subsection{Unique NE}
\begin{figure}
\centering
\includegraphics [height=10cm,width=8.5cm] {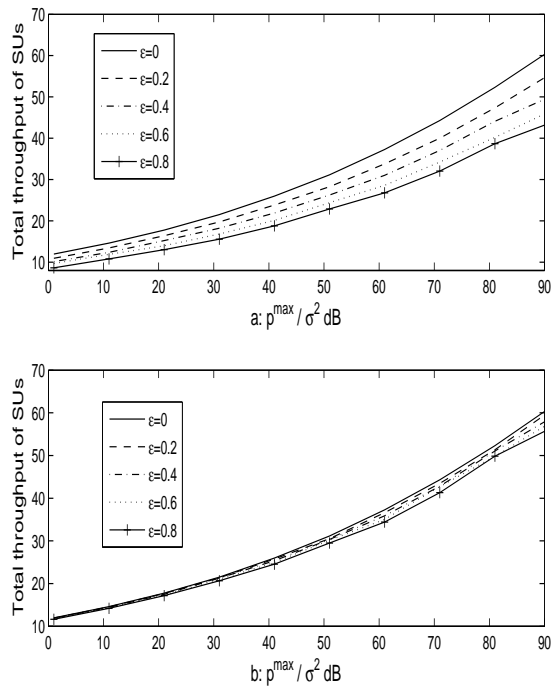}
\caption{Total throughput of SUs for different values of
$\varepsilon$ when Proposition 2 holds (i.e., when multiuser interference is low).}{\label{fig1}}
\end{figure}

Figs. \ref{fig1}(a) and \ref{fig1}(b) show the effect of
uncertainty on the total throughput of SUs when Proposition 2  holds, meaning that RNE is unique, and multiuser interference is low. In this set up, the number of SUs is 8, the number of sub-channels is 64, and
$p^{k}_{\scriptsize{\textnormal{mask}}}=p^{\scriptsize{\textnormal{max}}}$.
The values of $h^{k}_{ii}$, $h^{k}_{ji}$, and $\sigma^{k}_i$ are randomly chosen from the intervals $[0, 0.1]$, $[0, 0.01]$, and
$[0,0.01]$, respectively, guaranteing that Proposition 2 holds,
and are multiplied by fading coefficients. The estimated error is assumed to be symmetrically distributed in $[-\varepsilon,
\varepsilon]$ for all 64 sub-channels and for all SUs, and is added to the nominal value of $s_{i}^{k}$.

In Fig. \ref{fig1}(a), we assume that the value of $s_{i}^{k}$ is uncertain, but in Fig. \ref{fig1}(b), while the exact value of $s_{i}^{k}$ is available, it is assumed to be uncertain, i.e., $s_{i}^{k}$ is replaced by $(1+\varepsilon)s_{i}^{k}$. We take the average of total throughput values of all SUs for 20 realizations of channel gains, each with a different error value $\varepsilon$. Note that expanding the bounds on uncertainty, reduces the total throughput of SUs as compared to the case that there is no uncertainty, as expected form Lemma 1. The impact of uncertainty in Fig. \ref{fig1}(b) is much less than that in Fig. \ref{fig1}(a).

\begin{figure}
\centering
\includegraphics [height=10cm,width=8.5cm] {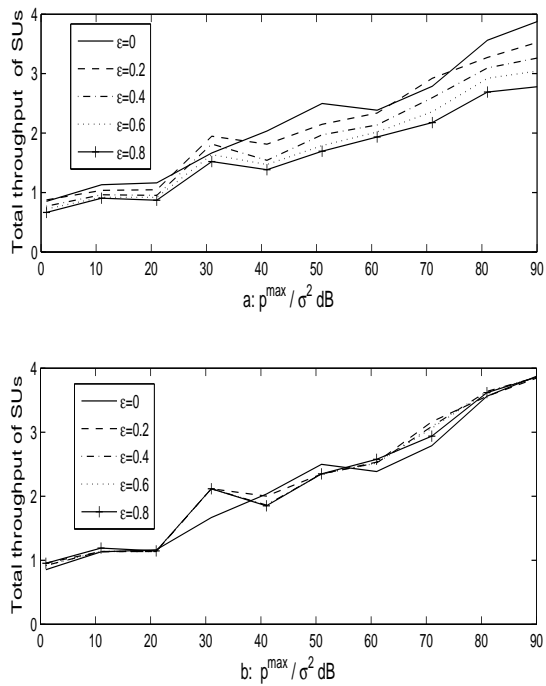}
\caption{Total throughput of SUs for different values of
$\varepsilon$ when Proposition 2 does not hold (i.e., when multiuser interference is high).}{\label{fig2}}
\end{figure}

\subsection{Multiple NEs}
To show the impact of uncertainty in cases that we encounter
multiple NEs, we assume that Proposition 2 does not hold, meaning that multiuser interference is very high. The values of $h^{k}_{ii}$, $h^{k}_{ji}$, and $\sigma^{k}_i$ are randomly chosen from the intervals $[0, 0.1]$, $[0, 1]$, and $[0,0.01]$ respectively. Again, we consider 8 SUs and 64 sub-channels, and take the average of total throughput values of all SUs for 20 realizations of channel gains, each with a different error value $\varepsilon$. The results are shown in Figs. \ref{fig2} (a) and (b). In Fig. \ref{fig2} (a), SUs encounter  uncertainty, and in Fig. \ref{fig2} (b), although the exact value of $s_{i}^{k}$ is available, it is assumed to be uncertain, i.e., $s_{i}^{k}$ is replaced by $(1+\varepsilon)s_{i}^{k}$.

As expected from Lemma 2, for some values of uncertainty, the robust game has a better performance than that of the conventional IWFA in both of the above cases, because uncertainty causes the game to converge to the orthogonal NE, resulting in a higher total throughput in some cases. %We proved this point, which
%is also briefly mentioned (but no proof is given) in
%\cite{Robusthaykin,ProbabilisticIWFA}.
The same is numerically shown in \cite{Robusthaykin}, where the
average throughput in the robust power allocation problem is shown to be very close to that of the conventional IWFA, and in
\cite{ProbabilisticIWFA}, where a higher throughput for the
probabilistic robust algorithm is reported as compared to that of the case where complete information is utilized by IWFA.

\begin{table}[h]
\caption{Channel Gains of 3 SUs} \vspace{-.2in}% title name of the table
\centering
\begin{tabular}{|c|c|c|c|c|c|c|}
  \hline
  k & 1 & 2 & 3 & 4& 5 & 6 \\
  \hline
  $h_{11}$ & 20.52 &  2.0  &  2.08 &  10.56  &  0.44 &   1.6 \\
$h_{12}$ &   4.91  &  4.97 &   3.95 &   3.94& 2.95& 5.95\\
$h_{13}$ &  7.9  &  5.97 &   2.97  &  4.92  &  1.93 & 6.94\\
$h_{21}$ & 0.92  & 0.94  &  0.95 &   0.92   & 0.95& 0.99\\
$h_{22}$ & 2.44&  26.32&   23.2&    3.64&    3.92& 0.68 \\
$h_{23}$ & 0.91  &  0.96   & 0.99   & 0.99  &  0.934 &0.95\\
$h_{31}$ & 0.91 &  0.95&    0.98&    0.98   & 0.93& 0.96\\
$h_{32}$ & 0.93&    0.96&    0.90&    0.96&    0.98&    0.97\\
$h_{33}$&  3.6&   24&    6&    1.6&   34& 40\\
$\sigma^{2}_{1}$ &2.2&    0.26&    4.1&    3.06&    0.02& 0.02\\
$\sigma^{2}_{2}$ &    8.24&    0.08&    0.18&    0.08& 0.04&0.06\\
  $\sigma^{2}_{3}$ &  0.22&    0.26&    4.08&    1.06&    0.02&
  0.02\\
  \hline
\end{tabular}
\label{tabel1}
\end{table}

In order to numerically investigate this effect, we consider 3 SUs, $K=6$, $p^{\scriptsize{\textnormal{max}}}=1$ and $p^{k}_{\scriptsize{\textnormal{mask}}}=0.5$ Watts, with channel gains and noise coefficients as in Table \ref{tabel1}. In Tables \ref{tabel2} and \ref{tabel3}, we show the allocated power to each SU in $\mathcal{G}$ and $\widetilde{\mathcal{G}}$, respectively. Note that at RNE of $\widetilde{\mathcal{G}}$, the allocated power levels to SUs are orthogonal to each other. Also, in the game $\widetilde{\mathcal{G}}$, the total aggregate throughput of SUs is increased, and in this special case, the throughput of each SU is also increased as compared to those of the game $\mathcal{G}$. However, in general, the throughput of each SU depends on its channel, and may vary.

\begin{table}[h]
\caption{Power Allocation by 3 SUs at NE of $\mathcal{G}$} \vspace{-.2in}% title name of the table
\centering
\begin{tabular}{|c|c|c|c|c|c|c|c|}
\hline
 $\text{SU}_i$ & $u_{i}$ & $p_1$ & $p_2$ & $p_3$ & $p_4$& $p_5$ & $p_6$\\
  \hline
  $\text{SU}_1$ &1.92&  0.44  &  0.1&     0 &   0.45 &  0  &       0  \\
$\text{SU}_2$& 3.82& 0   & 0.5&    0.5&     0  &       0 &        0 \\
$\text{SU}_3$ &  10.9 &0  &  0.0059    &0.3049  &   0  &  0.32 &   0.37 \\
\hline
\end{tabular}
\label{tabel2}
\end{table}

\begin{table}[h]
\caption{Power Allocation by 3 SUs at RNE of $\widetilde{\mathcal{G}}$ for $\varepsilon=3$} \vspace{-.2in}% title name of the table
\centering
\begin{tabular}{|c|c|c|c|c|c|c|c|}
  \hline
 $\text{SU}_{i}$ & $u_{i}$ & $p_1$ & $p_2$ & $p_3$ & $p_4$& $p_5$ & $p_6$  \\
  \hline
  $\text{SU}_1$ & 1.93 &  0.5&   0 & 0&    0.5&  0 &  0 \\
$\text{SU}_2$ & 3.95&0 &  0.5&    0.5&         0   &      0    &     0  \\
$\text{SU}_3$ &  11.17 & 0     &    0   &      0    &     0  &  0.5&    0.5 \\
\hline
\end{tabular}
\label{tabel3}
\end{table}

\begin{figure}
\centering
\includegraphics [height=7cm,width=9.5cm] {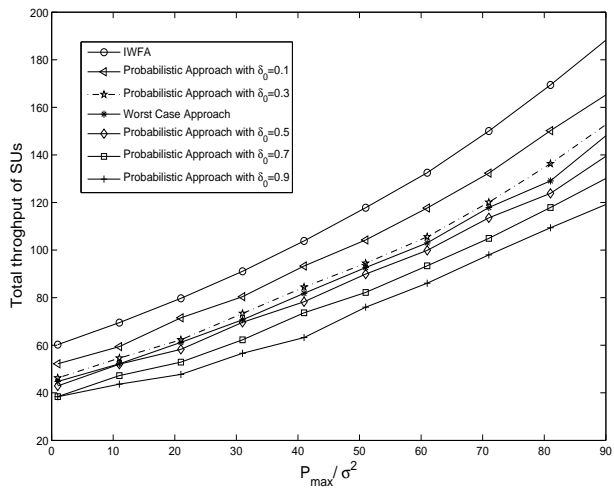}
\caption{Total throughput of SUs for different values of $\delta_0$ in the probabilistic approach as compared to that of the worst case approach when Proposition 2 holds (i.e., when multiuser interference is low).}{\label{withprobabilityuniquenash1}}
\end{figure}

\begin{figure}
\centering
\includegraphics [height=7cm,width=9.5cm] {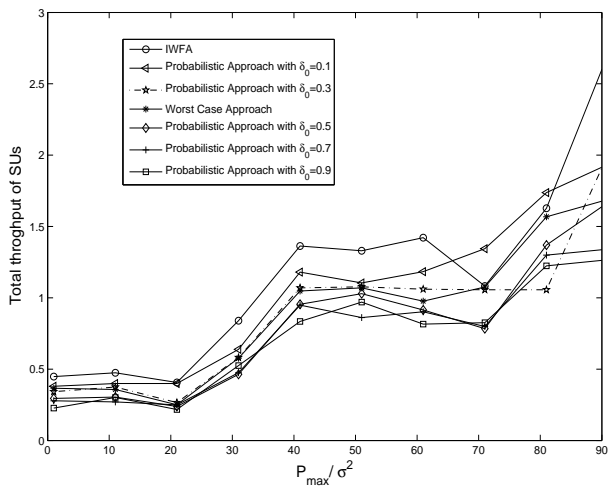}
\caption{Total throughput of SUs for different values of $\delta_0$ for the probabilistic approach as compared to that of the worst case approach when Proposition 2 does not hold (i.e., when multiuser interference is high).}{\label{withprobabilitymultiplenash3}}
\end{figure}

Finally, we compare the performance of our proposed RNE with that of a probabilistic approach in \cite{ProbabilisticIWFA}. In the probabilistic approach, by assuming that uncertainty of each parameter belongs to (\ref{Uncertainityregionlinear}) and its probability density function (pdf) is uniform, the SU's throughput in each subcarrier is guaranteed for a predefined probability $\delta_0$. Therefore, the uncertainty multiplier for $\bar{s}_{i}^{k}$ in
(\ref{utilitywithuncertainties}), i.e., $(1+\varepsilon)$ is
replaced by $(1-\varepsilon+2\times \varepsilon \times \delta_0)$. Obviously, when $0 < \delta_0 < 0.5$, the effect of uncertainty on NE of the probabilistic approach is less than that of the worst case approach; and when $0.5 \leq \delta_0 \leq 1$, the probabilistic approach is more robust than the worst case approach. This phenomenon is shown in Figs. \ref{withprobabilityuniquenash1} and \ref{withprobabilitymultiplenash3} for the case with a unique RNE and for multiple RNEs, respectively. In this simulation, for $\varepsilon=0.8$, the performances of the worst case approach and of the probabilistic approach are compared for different values of $\delta_0$. As we expected, when RNE is unique and $0 < \delta_0 < 0.5$, the probabilistic approach performs better than the worst case approach in terms of the achievable total throughput of SUs at RNE, and the opposite is true for $0.5 \leq \delta_0 \leq 1$. However, this cannot be extended to the case of multiple RNEs. As stated in Lemma 2, uncertainty may result in more orthogonality at RNE as compared to the non-robust approach and/or to the case with smaller values for uncertainty multipliers. In such cases, the total throughput may be higher at a RNE, as shown in Fig. \ref{withprobabilitymultiplenash3}.

\section{Conclusions}
We studied the impact of uncertainty in the channel state information and in interference levels on the total throughput of SUs in a spectrum sharing environment via robust game theory. To devise a robust power allocation in such environments, we focused on bounded and symmetrically distributed uncertainties, and showed that this game can be considered as a conventional IWFA with system parameters multiplied by bounds on uncertainty. The conditions for existence and uniqueness of RNE and for convergence of the proposed distributed algorithms of the robust game were derived based on the conventional IWFA. The performance of the robust game was compared to that of the conventional IWFA in terms of its total throughput. In the case of multiple RNEs, we showed that the orthogonal use of sub-channels by SUs at a RNE may lead to a higher total throughput of SUs as compared to that of the conventional IWFA.

\appendices
\section{Proof of Proposition 1}
As in (\!\!\cite{Robustgame}, Theorem 4), consider a robust finite game, with $M$ players, without private information, in which $U_{i}(\textbf{a}_{i}, \textbf{x}_{i})$ is a continuous and differentiable utility function for the $i^{{\scriptsize{\textnormal{th}}}}$ player, $\textbf{x}_{i}$ is the strategy of each player, $\textbf{a}_{i}$ is the vector of
uncertain parameters where $a_{ji}$ is its
$j^{{\scriptsize{\textnormal{th}}}}$ element in the following
interval
\begin{equation}
    a_{ji}=[a_{ji}^{\scriptsize{\textnormal{lower}}},a_{ji}^{\scriptsize{\textnormal{upper}}}], \nonumber
\end{equation}
where $a_{ji}^{\scriptsize{\textnormal{lower}}}$ and
$a_{ji}^{\scriptsize{\textnormal{upper}}}$ are the corresponding lower and upper bounds of uncertain parameters, respectively. In this game, $\{\textbf{x}_{1},\cdots,\textbf{x}_{M}\}$ is an equilibrium iff it is NE for the finite game with complete information, and with the same number of players and the same strategy space, where the utility function with deterministic parameters is
 \[ \widetilde{U}_{i}= \left\{
\begin{array}{l l}
 U(a_{ji}^{\scriptsize{\textnormal{lower}}}) \quad \text{if}  \quad \frac{\partial U_{i}(\textbf{a}_{i}, \textbf{x}_{i})}{\partial a_{ji}}\geq0\\
  U(a_{ji}^{\scriptsize{\textnormal{upper}}}) \quad \text{if} \quad  \frac{\partial U_{i}(\textbf{a}_{i}, \textbf{x}_{i})}{\partial a_{ji}}<0.
\end{array} \right. \]
In our robust game $\widetilde{\mathcal{G}}$, we have
%\begin{eqnarray}
  $\frac{\partial \widetilde{u}_{i}}{\partial s^{k}_i}<0$. %\nonumber
  %\end{eqnarray}
Therefore (\ref{utilitywithuncertainties}) follows from induction.

\section{Proof of Proposition 2}
For the complete information game $\mathcal{G}$, NE is unique \cite{Nash1} if
\begin{equation}\label{0.Proof of Proposition 2.}
\rho(\textbf{S}(k))<1 \quad \forall k \in \mathcal{K},
\end{equation}
where $\rho$ is the spectral radius of $\textbf{S}(k)$, and
$\textbf{S}(k)$ is a $M \times M$ matrix whose elements are
\begin{eqnarray}\label{1.Proof of Proposition 2.}
 S_{ij}(k) =  \left\{\begin{array}{l l}
0 \qquad\qquad \scriptsize{\textnormal{if}}\qquad i=j \\
 \frac{h^{k}_{ji}}{h^{k}_{ii}} \qquad \;\quad\;\!\! \scriptsize{\textnormal{if}} \qquad i\neq j. \end{array} \right.
 \end{eqnarray}
From (\ref{Uncertainityregionlinear}), the robust game
$\widetilde{\mathcal{G}}$ has a unique
RNE if
\begin{equation}\label{2.Proof of Proposition 2.}
    \max_{s\in \mathcal{R}_{s}} \rho (\textbf{S}(k))<1.
\end{equation}
In this case, we have
%\begin{equation}\label{3.Proof of Proposition 2.}
   $\textbf{S}(k)= \bar{\textbf{S}}(k)+ \hat{\textbf{S}}(k)$,
%\end{equation}
where $\bar{\textbf{S}}(k)$ is a $M \times M$ matrix whose elements are \begin{eqnarray}\label{4.Proof of Proposition 2}
 \bar{S}_{ij}(k) =  \left\{\begin{array}{l l}
0 \qquad\qquad \scriptsize{\textnormal{if}}\qquad i=j \\
 \frac{\bar{h}^{k}_{ji}}{\bar{h}^{k}_{ii}} \qquad \;\quad\;\!\! \scriptsize{\textnormal{if}} \qquad i\neq j, \end{array} \right.
 \end{eqnarray}
and $\hat{\textbf{S}}(k)$ is a matrix of the uncertain parts of
system parameters in which the sum of its
$i^{{\scriptsize{\textnormal{th}}}}$ row is less than
 $\varepsilon^{k}_{i}$. Recall that Frobenius norm of a matrix is  $\|\bar{S}_{ij}(k)\|_{F}=\sum_{i,j}\bar{S}^{2}_{ij}(k)$ and that  $\|\bar{S}_{ij}(k)\|_{2}\leq\|\bar{S}_{ij}(k)\|_{F}$. Thus
 \begin{eqnarray}\label{5. Proof of Proposition 2}
     \max_{s\in \mathcal{R}_{s}} \rho (\textbf{S}(k))\leq  \|\bar{S}_{ij}(k)\|_{2}+   \|\hat{S}_{ij}(k)\|_{2} \leq \|\bar{S}_{ij}(k)\|_{2}+
     \|\hat{S}_{ij}(k)\|_{F}\leq\|\bar{S}_{ij}(k)\|_{2}+\|\mathbf{s}^{k}\|_2.
\end{eqnarray}
On the other hand, from (\!\!\cite{Matrixanalysis1}, Theorem 1.1)
\begin{equation}\label{6.Proof of Proposition 2}
 \max_{_{\|\textbf{G}\|_{F} < 1}} \rho (\textbf{F}+\textbf{G})\leq  \rho
 (\frac{\textbf{F}+\textbf{F}^{T}}{2})+1. \nonumber
\end{equation}
Thus
\begin{equation}\label{7.Proof of Proposition 2}
 \max_{_{\|\hat{\textbf{S}}\|_{F} < \|\mathbf{s}^{k}\|_2}} \rho (\bar{\textbf{S}}+\hat{\textbf{S}})\leq  \rho
 (\frac{\bar{\textbf{S}}+\bar{\textbf{S}}^{T}}{2})+\|\mathbf{s}^{k}\|_2. \nonumber
\end{equation}
For a symmetric $\textbf{S}$, we have $\bar{\textbf{S}}=\bar{\textbf{S}}^{T}$,
$\rho(\frac{\bar{\textbf{S}}+\bar{\textbf{S}}^{T}}{2})=\rho(\bar{\textbf{S}})\leq\|\bar{S}_{ij}(k)\|_{2}$,
and therefore, (\ref{Proposition 2.1}) reduces to (\ref{Proposition 2.2}).

\section{Proof of Lemma 1}
To prove this lemma, we use variational inequalities (VI) to reformulate NE, and perform sensitivity analysis on VI to derive the relationship between $\mathbf{p}^*$ and $\widetilde{\mathbf{p}}^*$.

Step 1: Consider that the nominal NE is the solution to the following variational inequalities (Proposition 1.4.2 in \cite{PangVI} and \cite{VIintroduction}),
\begin{equation}\label{RNEVI}
    \mathbf{p}^*= \text{Solution to}(VI(\mathcal{P}, \mathcal{F}))
\end{equation}
where $\mathcal{P}=\prod_{i \in \mathcal{M}}\mathcal{P}_i$, $\mathcal{F}=(\mathcal{F}_i)$, $\mathcal{F}_i= \boldsymbol{\sigma}_i+\sum_{j=1}^{M}\mathbf{M}_{ij}\mathbf{p}_{j}$, $\boldsymbol{\sigma}_i=(\sigma_{i}^{k}/h_{ii}^{k})_{k=1}^{K}$, and $\mathbf{M}_{ij}=\text{diag}(\frac{h_{ij}^{k}}{h_{ii}^{k}})_{j=1}^{M}$. Considering uncertainty in the parameters can be viewed as perturbation in $\mathcal{F}=(\mathcal{F}_i)$, which we show by $\widetilde{\mathcal{F}}=(\widetilde{\mathcal{F}}_i)$, where $\widetilde{\mathcal{F}}_i$ has the same definition as $\mathcal{F}_i$ except that $\widetilde{\boldsymbol{\sigma}}_i=(\sigma_{i}^{k}(1+\epsilon_{i}^{k})/h_{ii}^{k})_{k=1}^{K}$ and $\mathbf{M}_{ij}=\text{diag}(\frac{h_{ij}^{k}(1+\epsilon_{i}^{k})}{h_{ii}^{k}})_{j=1}^{M}$. Therefore we consider the solution to $\widetilde{\mathcal{G}}$ as a solution to $VI(\mathcal{P}, \mathcal{F}+q)$, where $q=\max \epsilon_{i}^{k}\frac{h_{ij}^{k}}{h_{ii}^{k}}$.

Step 2:  When Proposition 1 holds, $F$ is strongly monotone.

Step 3: When $F$ is strongly monotone, the solution to $VI(\mathcal{P}, \mathcal{F}+q)$, i.e., $\varphi(q)$ is a monotone plus single-valued map (2.9.17 in \cite{PangVI}). Recall that the solution to the power control game, denoted by $\varphi(q)$, is a decreasing function of $q$, i.e., %\begin{equation}\label{singlemonotobemap}
    if $q_1 < q_2 \Longrightarrow \varphi(q_1) \geq \varphi(q_2)$.
%\end{equation}
Hence, by increasing uncertainty, the power allocated to each user is reduced as compared to that of the nominal game. Increasing uncertainty also reduces the social utility, because the utility function in Proposition 1 is strictly convex.

\section{Proof of Lemma 2}

Assume that the $i^\text{th}$ SU converges to $\mathbf{p}_{i}^{*}$ at NE of
$\mathcal{G}$, for which there is a corresponding $s_{i}^{k*}$, and the set of all sub-channels with nonzero power allocation by the $i^{\scriptsize{\textnormal{th}}}$ SU is denoted by $\mathcal{I}_{i}^{*}$. Consequently,
$\widetilde{\mathbf{p}}_{i}^{*}$,
$\widetilde{s}_{i}^{*k}$, and
$\widetilde{\mathcal{I}}_{i}^{*}$ belong to the game
$\widetilde{\mathcal{G}}$. In what follows, we will show that
$\widetilde{\mathcal{I}}_{i}^{*}\subseteq \mathcal{I}_{i}^{*}$. In doing so, we denote the Lagrange multipliers at NE of the games
$\mathcal{G}$ and $\widetilde{\mathcal{G}}$ by $\lambda_{i}^{*}$ and $\widetilde{\lambda}_{i}^{*}$, respectively, which are
increasing functions of $s_{i}^{k}$. Therefore, for the $i^\text{th}$ SU we have
\begin{eqnarray}  \label{Lemma 2.1}
 \frac{1}{\widetilde{\lambda}_{i}^{*}}\leq \frac{1}{\lambda_{i}^{*}}= \sum_{k \in \mathcal{I}_{i}^{*} }
s_{i}^{k*}+p_{i}^{\scriptsize{\textnormal{max}}} <
\frac{\sum_{i\neq j} p^{q*}_{j}h^{q}_{ji}+\sigma^{q}}{h_{ii}^{q}},
\end{eqnarray}
where $q \notin \widetilde{\mathcal{I}}_{i}$. Obviously, $\mathbf{p}_{i}^{*}$ of the $i^\text{th}$ SU leads to $s^{k*}_{i}\leq s^{k}_{i} \quad \forall i , k$. Therefore, for any power allocation strategy $\mathbf{p}_{i} \in \mathbf{p}$, we have
\begin{eqnarray}\label{Lemma 2.2}
\frac{\sum_{i\neq j} p^{q*}_{j}h^{q}_{ji}+\sigma^{q}}{h_{ii}^{q}}
<(1+\varepsilon^{q}_{i})\frac{\sum_{i\neq
j}p^{q}_{j}h^{q}_{ji}+\sigma^{q}}{h_{ii}^{q}}.
\end{eqnarray}
From (\ref{Lemma 2.1}) and (\ref{Lemma 2.2}), those sub-channels that are not used in $\mathcal{G}$ are not used in
$\widetilde{\mathcal{G}}$ as well. Now, as stated in (\!\!\cite{Nash1}, Proposition 3), when
multi-user interference is high, an orthogonal NE always exists
for the game $\mathcal{G}$. When the $i^{\scriptsize{\textnormal{th}}}$ SU chooses a $q' \in
\widetilde{\mathcal{I}}_{i}$, it is orthogonal to those of other users. Assuming this, for any other power allocation profile of other users, we have
\begin{equation}\label{Lemma 2.3}
(1+\varepsilon^{q'}_{i})\frac{\sigma^{q'}}{h_{ii}^{q'}}<
(1+\varepsilon^{q'}_{i})\frac{\sum_{i\neq j}
p^{q'}_{j}h^{q'}_{ji}+\sigma^{q'}}{h_{ii}^{q'}}, \quad \forall i \in \mathcal{M}.
\end{equation}
Since $\widetilde{s}_{i}^{*k}\gg 1$, the utility at RNE for each SU is higher than that of any other strategy profile. Hence, there is no incentive for other users to change their strategy profile from the orthogonal RNE. As such, RNE of $\widetilde{\mathcal{G}}$ is more orthogonal than NE of $\mathcal{G}$.
\section{Proof of Proposition 3}
For the game with complete information $\mathcal{G}$, it is proved in \cite{Nash2} that both sequential and simultaneous distributed algorithms converge if
\begin{equation}\label{1. Proof of Proposition 3.}
\rho(\textbf{S}^{\scriptsize{\textnormal{max}}})<1 \quad \forall k \in \mathcal{K},
\end{equation}
where $\rho$ is the spectral radius of
$\textbf{S}^{\scriptsize{\textnormal{max}}}$, and $\textbf{S}^{\scriptsize{\textnormal{max}}}$ is a $M \times M$ matrix whose elements are
\begin{eqnarray}\label{2. Proof of Proposition 3.}
 S_{ij}^{\scriptsize{\textnormal{max}}}=  \left\{\begin{array}{l l}
0 \qquad\qquad \qquad\,\;\scriptsize{\textnormal{if}}\qquad i=j \\
\max_{k \in \mathcal{K}} \frac{h^{k}_{ji}}{h^{k}_{ii}} \qquad
\scriptsize{\textnormal{if}} \qquad i\neq j.
\end{array} \right.
 \end{eqnarray}
From (\ref{Uncertainityregionlinear}), the robust game
$\widetilde{\mathcal{G}}$ converges to RNE by using the iterative sequential and simultaneous distributed algorithms if
\begin{equation}\label{3. Proof of Proposition 3.}
    \max_{s_{i}^{k} \in \mathcal{R}_{s} , \forall i, k} \rho (\textbf{S}^{\scriptsize{\textnormal{max}}})<1.
\end{equation}
In this case, we have $\textbf{S}^{\scriptsize{\textnormal{max}}}= \bar{\textbf{S}}^{\scriptsize{\textnormal{max}}}+
   \hat{\textbf{S}}^{\scriptsize{\textnormal{max}}}$, where $\hat{\textbf{S}}^{\scriptsize{\textnormal{max}}}$ is the uncertain parts of $\textbf{S}^{\scriptsize{\textnormal{max}}}$, whose elements are
\begin{eqnarray}\label{5. Proof of Proposition 3.}
 \bar{S}_{ij}^{\scriptsize{\textnormal{max}}}=  \left\{\begin{array}{l l}
0 \qquad\qquad \;\;\;\scriptsize{\textnormal{if}}\qquad i=j \\
\max \frac{\bar{h}^{k}_{ji}}{\bar{h}^{k}_{ii}} \qquad
\scriptsize{\textnormal{if}} \qquad i\neq j. \end{array} \right.
 \end{eqnarray}
Since $\hat{S}_{ij}^{\scriptsize{\textnormal{max}}}\leq \max
\varepsilon_{i}^{k} \bar{s}_{i}^{k}$, from (\ref{6.Proof of Proposition 2}) we have
\begin{eqnarray}\label{6.Proof of Proposition 3}
      \max_{s\in \mathcal{R}_{s}} \rho (\textbf{S}^{\scriptsize{\textnormal{max}}})\leq  \|\bar{S}_{ij}^{\scriptsize{\textnormal{max}}} \|_{2}+ \|\hat{S}_{ij}(k)\|_{F}\leq  \|\bar{S}_{ij}^{\scriptsize{\textnormal{max}}}\|_{2}+ \sqrt{|\mathcal{M}|}\|\mathbf{s}^{\scriptsize{\textnormal{max}}}\|_{2}.
\end{eqnarray}

%\IEEEtriggeratref{2}
\bibliographystyle{IEEEtran}
\bibliography{IEEEabrv,mybib}
\end{document}